# FMCW SAR with New Synthesis Method Based on A-SPC Technique

Junhyeong Park, *Member, IEEE*, Dae-Hwan Jung, and Seong-Ook Park, *Senior Member, IEEE*

*Abstract*—Frequency modulated continuous wave (FMCW) radar is emerging as a trendy radar system for synthetic aperture radar (SAR). This letter proposes a novel method for the extraction of the SAR image with the FMCW radar. The proposed method can improve the quality of the SAR image. For the verification, we built an automobile SAR (AutoSAR) system and conducted experiments to extract the SAR map by using the AutoSAR system. Then, we synthesized SAR images through both the conventional method and the proposed method to demonstrate the performance of the proposed method. The experimental results show that the SAR image has been successfully improved by the proposed method.

## I. Introduction

FREQUENCY modulated continuous wave (FMCW) radars have been frequently used as a synthetic aperture radar (SAR) in recent years due to various advantages [1]–[5]. The FMCW SAR has not only a cost-effective high-resolution sensing ability but also the property of lightweight and small-sized hardware. However, the leakage between the transmitter (TX) and the receiver (RX), which is an inherent problem of the FMCW radar, considerably decreases the sensitivity of the radar [6]–[9].

In earlier studies, we have proposed the stationary point concentration (SPC) technique that was a new technique for the leakage mitigation in the FMCW radar [6]–[8], and we have recently proposed the advanced SPC (A-SPC) technique that overcomes the limitations of the SPC technique [9]. However, the SPC and the A-SPC techniques have been developed for the small drone detection that can be assumed to be the detection of discrete point targets. The promising possibility that these techniques can also be used to improve the results of more advanced radar applications such as SAR or inverse SAR than the simple detection could be found in [8], [9]. Nevertheless, no proposals have been made so far and have not been verified.

In this letter, we verify for the first time that the A-SPC technique can be used for SAR, an advanced radar application that detects and synthesizes continuous spatial target signals.

We propose a novel SAR image synthesis method based on the A-SPC technique. The proposed method improves the quality of the SAR image by reducing the noises in the SAR image. In addition, since the proposed method can mitigate the leakage, it reduces the burden of physically increasing the distance between the TX and the RX to attenuate the leakage. Thus, the proposed method can help to reduce the size of the SAR platform.

To show the performance of the proposed method, we built an automobile SAR (AutoSAR) system which installs the SAR on a van. We carried out experiments by scanning the targeted area with the self-developed AutoSAR. Then, we extracted each resulting SAR map through the conventional method and the proposed method. Finally, the comparison results demonstrate that the proposed method successfully improves the SAR image.

## II. Proposed Method

Fig. 1 shows the proposed method as a form of flow chart. Note that $n$ is the index for the fast time domain, and $m$ is the index for the slow time domain. The proposed method continually applies the A-SPC technique to complex raw data, $x[n,m]$, for every $m$th bunch of beat signals until $m$ reaches $M$ that is the index for the last bunch of beat signals to synthesize the SAR image. The A-SPC technique included in the proposed method finds the frequency, $f_{IF\ beat\ leakage}(m)$, and the phase, $\theta_{IF\ leakage}(m)$, of the leakage signal according to the formulas in Fig. 1, where $k_{IF\ leakage}(m)$ is the index number for the leakage signal, $X[k,m]$ is the result of the $NFFT$-point fast Fourier transform (FFT) of $x[n,m]$ along the fast time domain. $F_S$ is the sampling frequency, $NFFT$ is the total number of data samples and zero-pads for the zero-padding, and $\angle X$ is the phase response of $X[k,m]$. With the found $f_{IF\ beat\ leakage}(m)$ and $\theta_{IF\ leakage}(m)$, a digital numerically controlled oscillator, $NCO[n,m]$, is generated, then the complex-based mixing is carried out by taking the conjugate to $NCO[n,m]$ and multiplying it with $x[n,m]$. The real data in the result of the complex-based mixing are extracted as the output of the A-SPC technique.

These procedures leads to the leakage mitigation by concentrating the phase noise of the leakage signal on the stationary point in the sinusoidal function of the leakage signal [9]. Thus, the signal-to-noise ratio (SNR) is improved as the noise floor that is dominated by the phase noise of the leakage signal is reduced. Moreover, the internal delay compensation and the phase calibration of the system, which are the additional functions of the A-SPC technique, can be obtained [9]. Because the proposed method iterates the procedures of the A-SPC technique for every $m$th bunch of beat signals, the SNR in the slow time domain as well as the SNR in the fast time domain can be naturally improved. Hence, the quality improvement in the synthesized SAR image can be expected. After the iteration of the A-SPC technique and the data arrangement, the proposed method applies the FFT along the fast time domain for the range compression as shown in Fig. 1 and finish with the common procedures in the range-Doppler algorithm for SAR.

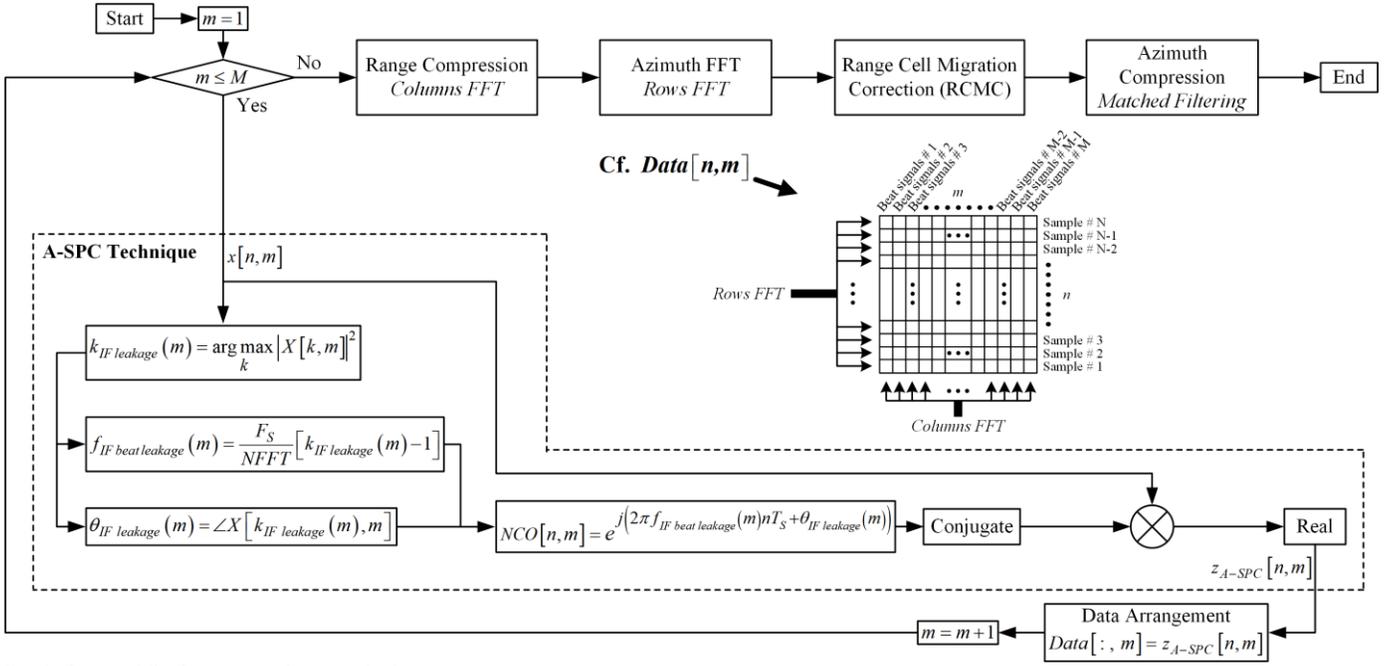

Fig. 1. Proposed SAR image synthesis method.

TABLE I
SPECIFICATIONS AND PARAMETERS OF THE FMCW SAR

| Parameters | Values |
| --- | --- |
| Radar configuration | Quasi-monostatic |
| System architecture | Heterodyne |
| Operating frequency | 14.35-14.50 GHz |
| Transmit power | 30 dBm |
| Antenna | Corrugated horn |
| Antenna gain | 16 dBi |
| Antenna half power beamwidth | 34° |
| Sweep bandwidth ($BW$) | 150 MHz |
| True range resolution | 1 m |
| Final IF carrier frequency ($f_{IF\,carrier}$) | 0 MHz |
| NFFT for finding out $f_{IF\,beat\,leakage}$ and $\theta_{IF\,leakage}$ | $2^{19}$ |
| Sweep period ($T$) | 800 us |
| Desired digital bandwidth | 2.5 MHz |
| Sampling frequency ($F_S$) | 5 MHz |
| Maximum unambiguous range | 2000 m |
| Window function | Hann |
| Platform (vehicle) speed | 60 km/h |

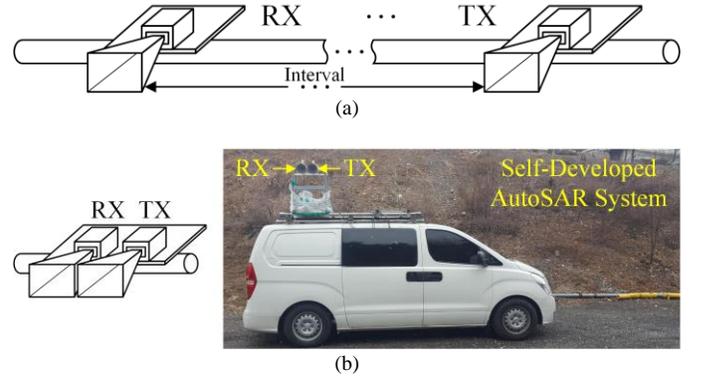

Fig. 2. Comparison of the FMCW SAR platforms. (a) Common FMCW SAR platform. (b) FMCW SAR platform when the proposed method is considered to be applied, and a photo of the self-developed AutoSAR system.

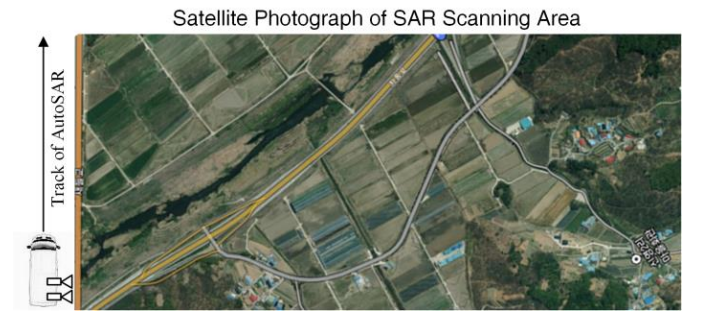

Fig. 3. Experimental scenario.

## III. EXPERIMENTS, RESULTS, AND DISCUSSION

For the verification of the proposed method, we built an AutoSAR system with *Ku*-band FMCW radar hardware in [6]. The specifications and parameters of the AutoSAR are listed in Table I. Fig. 2 shows the comparison of the FMCW SAR platforms. In the common FMCW SAR platform, there is an interval between the TX and the RX to mitigate the leakage [1]–[5]. On the other hand, the interval can be greatly decreased if the proposed method is applied for the FMCW SAR, because the proposed method can mitigate the leakage. Therefore, the size of the platform can be reduced as shown in Fig. 2(b). Fig. 3 shows the experimental scenario. We drove the AutoSAR on a bridge and scanned rural villages. Then, we synthesized each SAR image with the conventional method and the proposed method, respectively.

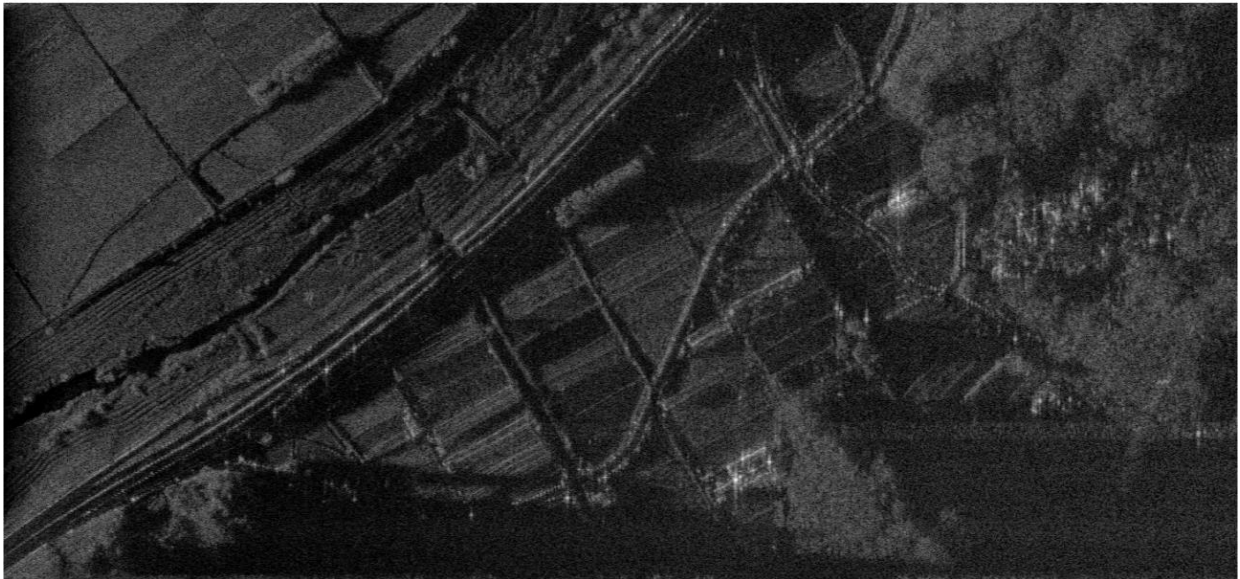

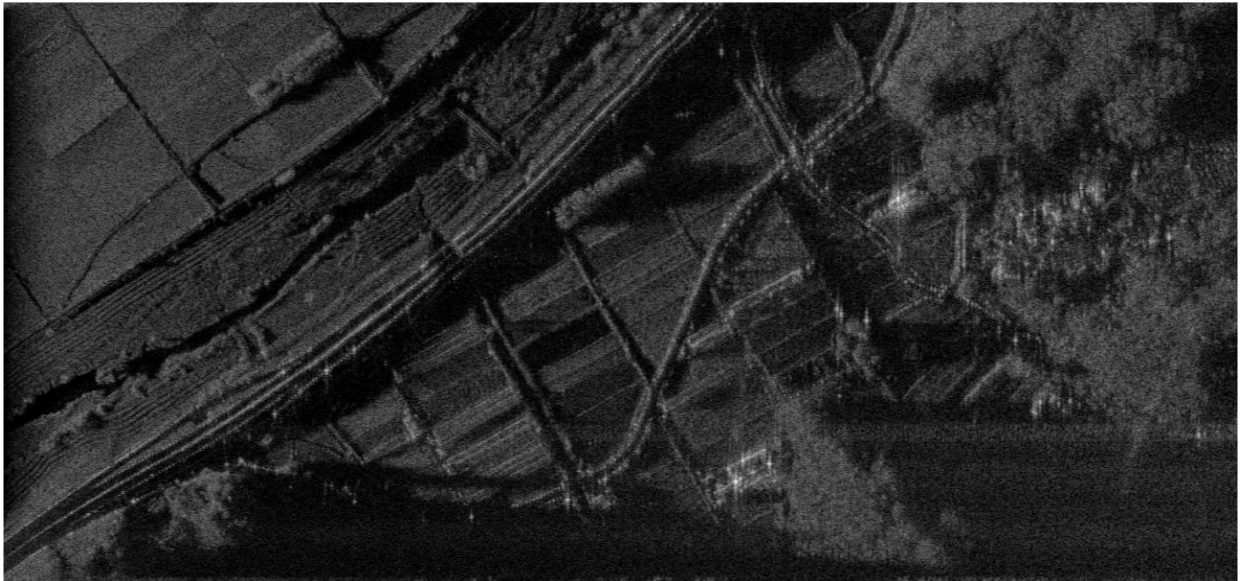

Fig. 4. Extracted SAR images. (a) Resulting SAR image through the conventional method. (b) Resulting SAR image through the proposed method.

The resulting SAR images whose area covers about 1200 m × 583 m are displayed in Fig. 4. The SAR image obtained through the proposed method is more clear and cleaner than the SAR image obtained through the conventional method as the speckle noises have been significantly reduced. Therefore, it has been verified that the proposed method improves the quality of the SAR image.

## IV. Conclusion

A novel FMCW SAR image synthesis method based on the A-SPC technique has been proposed. Since the proposed method includes the A-SPC technique that mitigates the leakage, we were able to minimize the size of the FMCW SAR platform by narrowing the interval between the TX and the RX as much as possible. As a result of synthesizing the SAR image through the proposed method after scanning the rural villages with the self-developed FMCW AutoSAR, we have confirmed that the proposed method has successfully improved the SAR image. Therefore, it has been demonstrated that the proposed method is useful and practical for FMCW SAR.